\documentclass[10pt,twocolumn,pre,aps,superscriptaddres,floatfix]{revtex4-1}

\usepackage{amssymb}
\usepackage{amsmath}
\usepackage{bm}

\usepackage[linkcolor = blue, citecolor = blue, urlcolor = blue, colorlinks = true]{hyperref}

\usepackage{graphicx}
\usepackage{wrapfig}
\usepackage[dvipsnames]{xcolor}

\usepackage[utf8]{inputenc}
\usepackage{lipsum}

\begin{document}

\title{Delayed Gravitational Collapse of Attractive Colloidal Suspensions}

\author{K.W. Torre}
 \email{k.w.torre@uu.nl}
\affiliation{
 Institute for Theoretical Physics, Center for Extreme Matter and Emergent Phenomena, Utrecht University, Princetonplein 5, 3584 CC Utrecht, The Netherlands 
}

\author{J. de Graaf}
\affiliation{
 Institute for Theoretical Physics, Center for Extreme Matter and Emergent Phenomena, Utrecht University, Princetonplein 5, 3584 CC Utrecht, The Netherlands 
}

\date{\today}


\begin{abstract}
Colloidal gels have strong industrial relevance as they can behave liquid- and solid-like. The latter allows them to support the buoyant weight against gravity. However, the system is intrinsically out-of-equilibrium, which means that the colloids must eventually settle out of the suspension. The process of settling has been captured theoretically, but the presence of a delay time during which the gel appears relatively unaffected by gravity has not. Here, we modify existing frameworks to capture this delay, by treating the gel as a continuum with viscoelastic response that is based on the local bond density. We can solve our model numerically to obtain the evolution of the colloid density profile and recover qualitatively the accumulation of a dense layer on top of the settling gel, as is experimentally observed in depletion gels. This numerical study is complemented by a theoretical analysis that allows us to identify an emergent time and length scale that set the dynamics of the gel. Our model provides a solid foundation for future studies that incorporate hydrodynamic erosion and tackle industrially relevant geometries.
\end{abstract}

\maketitle


\section{\label{sec:intro}Introduction}

Colloidal gels, formed through the aggregation of micron-sized particles within a solvent medium~\cite{lekkerkerker1992poon, poon2002physics, bergenholtz2003gelation, chen2004microscopic}, represent a class of soft materials with diverse applications ranging from food and care products~\cite{larson1999structure,food-soft-materials2005}, to crop protection formulations~\cite{crop_protection2006}, and bio-fluids~\cite{darras2022}. In any real-world system, a density mismatch between the solvent and colloidal particles is unavoidable~\cite{buscall1987consolidation}. This leads buoyancy forces and subsequent sedimentation (or creaming) of the colloids, ultimately giving rise to a separation of the suspension into colloid-rich and colloid-poor regions. One of the appealing aspects of using colloidal gels is that they possess the ability to support the particles' buoyant weight against gravity for a finite, often extended, duration~\cite{allain1995aggregation, allain2001systematic, starrs2002collapse, allain2003rapid, weitz2005gravitational, bartlett2012sudden, zaccarelliHarich2016, padmanabhan2018gravitational}. This time is typically referred to as a delay time and sets the typical shelf life of a gel-based product in many applied contexts~\cite{zaccarelliHarich2016,zhou2018onset}.

The gel's ability to support its own weight can be attributed to its network-like structure~\cite{swan-furst2019} and the internal dynamics~\cite{zaccarelli2009colloidal}.  A colloidal gel forms due to the interplay between thermal/Brownian diffusion and strong, short-ranged attractions between the colloids. The latter cause the colloids to aggregate into a network structure, when the system is quenched into the spinodal region of the phase diagram~\cite{carpineti1992spinodal}. For attractions that are several times the thermal energy --- $k_{\mathrm{B}}T$ with $k_{\mathrm{B}}$ Boltzmann's constant and $T$ the temperature --- rearrangement of the formed network is slowed down~\cite{foffi2002evidence}. This significantly extends the time it takes the system to relax to thermodynamic equilibrium~\cite{zaccarelli2007colloidal, Royall2021}. The arrested dynamics also halts~\cite{zaccarelliHarich2016,darras2022} or, at the very least, strongly slows down~\cite{allain2003rapid, weitz2005gravitational} buoyancy-mediated separation. The amount by which this separation is suppressed, depends sensitively on the way the gel is prepared and the interactions between the colloids. Consequently, the time scales reported for complete collapse can vary significantly, ranging from minutes to months~\cite{zaccarelliHarich2016}.

Accurately characterizing a gel's resistance to gravity in a physical model presents a considerable challenge. This is because experiments reveal that colloidal gels can exhibit several settling behaviors: fast sedimentation, delayed rapid collapse, and slow sedimentation~\cite{zaccarelliHarich2016}. Which of these is present in a given suspension, primarily dependents on the initial volume fraction $\phi_0$ of the sample, the strength of attractions between colloids $\epsilon_0$, and the size of the particles~\cite{allain2001systematic, allain2003rapid, weitz2005gravitational, zaccarelliHarich2016, darras2022, buzzaccaro2012equilibrium}. The importance of size can be understood by considering the gravitational P{\'e}clet number, which gives the dimensionless ratio between sedimentation and diffusion. For a sphere of radius $a$, the number is given by
\begin{align}
\label{eq:Pegrav} \mathrm{Pe}_{g} &= \frac{4 \pi g \Delta \rho a^{4}}{3 k_{\mathrm{B}} T} ,
\end{align}
where $g$ is the acceleration of gravity and $\Delta \rho$ is the density difference between the colloid and suspending fluid. The strong dependence on size follows from $\mathrm{Pe}_{g} \propto a^{4}$.

Harrich~\textit{et al.}~\cite{zaccarelliHarich2016} experimentally investigated the settling dynamics of depletion gels comprising $\sim 0.65$~$\mu$m colloids. They established a $\phi_{0}-\epsilon_{0}$ state diagram for the types of behavior that were observed. The most striking of these is delayed gravitational collapse~\cite{zaccarelliHarich2016, zhou2018onset} that occurs for $0.15 \lesssim \phi_{0} \lesssim 0.35$. In these situations, the interface between the gel and supernatant phase --- a region (nearly) devoid of particles --- initially resists sedimentation. However, at a certain point, the system abruptly looses structural integrity and enters a regime of rapid settling, which is characterized by a constant velocity of the interface. This regime concludes when the interface reaches the densified bottom part of the sample, after which the velocity of the interface exponentially decays to zero as the colloid-rich phase compacts. It was suggested in Ref.~\cite{zaccarelliHarich2016} and later shown in Ref.~\cite{zhou2018onset}, that the accumulation of dense debris on top of the gel triggers the catastrophic failure. This debris originates from the curved parts of the meniscus between air and the sample.

Several attempts have been made to theoretically explain the response of colloidal gels to gravitational forces. In 2003, Derec~\textit{et al.}~\cite{allain2003rapid} studied experimentally the rapid collapse of gels formed from strongly aggregating colloidal suspensions, for low values of the initial volume fraction ($\phi_0<1$\%). They showed that in the first resistant regime, the gel-supernatant interface slowly settles at a constant velocity. It was later argued that this follows a power law $\phi_0^{(1-D)/(3-D)}$, with $D$ the fractal dimension of the gel network~\cite{allain1995aggregation}. In addition, Ref.~\cite{allain2003rapid} connects the transition into the second, linear regime to the appearance of fractures within the bulk of the gel. These provide an easy route for the solvent to move up into the supernatant phase, which explains the abrupt increase in the interface's settling velocity. The authors also proposed a simple hydrodynamic model by which the increase in velocity can be estimated, which takes as input the height and radius of a typical crack in the gel.

These insights were built upon in the work of Manley~\textit{et al.}~\cite{weitz2005gravitational}, wherein the collapse dynamics is determined by the balance between gravitational stress and the yield stress of the network ($\phi_0<1$\%). When the former is larger, a gel collapses poroelastically, with a rate of compression that decays exponentially in time. Otherwise, the network eventually yields, leading to rapid linear settling. The authors use a modified Carman-Kozeny relation~\cite{happel1991low} to argue that the characteristic pore size in the gel network is set by the largest length scale in the system, the cluster size. They also show that their experimental data collapses onto a single curve that is well-described by their model. However, it is important to note that both studies~\cite{allain2003rapid, weitz2005gravitational} focus primarily on colloidal particles with small diameters, typically in the range of tens of nanometers, which can explain some of the differences to the observations in Refs.~\cite{zaccarelliHarich2016,zhou2018onset}.

More recently, Darras~\textit{et al.}~\cite{darras2022} investigated the sedimentation behavior of dense soft colloidal gels, formed by the aggregation of red blood cells ($\phi_{0} \gtrsim 0.25$). Unlike previous models that assumed low initial volume fractions to simplify hydrodynamic descriptions, they developed a theoretical framework capable of accurately reproducing the time evolution of the gel interface height. Their model operates on the assumption that the gel network uniformly compresses as it sediments, resulting in a mostly homogeneous colloid distribution over time. This is a reasonable approximation for the dense and weak gels they examined. However, while their model successfully captures many aspects of the gel sedimentation process, the initial delay time observed in their gels does not arise spontaneously from the theory and is instead introduced as an adjustable parameter.

Motivated by the absence of theoretical descriptions that capture a delay time, we present here a comprehensive theoretical framework that overcomes this issue. We also account for spatial variation of the gel properties, which should allow us to shed light on phenomena like debris formation. In our model, we describe the stress within the gel as a dilatational viscous response to applied compression~\cite{landau1959fluid}. Following other analyses, Darcy's law is used to model the dynamic coupling between the solvent and colloidal phases~\cite{carman1939permeability, gray2004examination, carrillo2019darcy}. The average network pore cross-section is determined by the average surface-to-surface distance between particles in our approach. We show that these elements are sufficient to qualitatively capture features observed in experiments, including the emergence of delay times and the accumulation of debris at the top of the gel.

The remainder of this paper is organized as follows. In Section 2, we present the theoretical framework in generic dimensions and boundary conditions. In Section 3, we first provide the analytical solution for the instantaneous mean colloidal velocity in samples with homogeneous volume fractions. Subsequently, numerical results depicting the time evolution of the gel-supernatant interface, as well as colloidal density profiles, are presented. We conclude the section by demonstrating how the delay time can be accurately predicted from the characteristics of colloidal flow at the beginning of the collapse. In Section 4, we delve into a detailed discussion of the obtained results, addressing the limitations of the models and establishing connections with experimental observations. Finally, in Section 5, we draw conclusions and explore potential future directions, including the study of meniscus curvature on the behavior of a gel.

\section{\label{sec:theory}Theoretical Model}

In this section, we provide a step-by-step derivation of our model. As the primary focus of this paper is the theoretical result, we have included details on the numerical solving of the model in Appendix~\ref{sec:appen}.

\subsection{\label{sec:mass}Mass Conservation Equation}

Consider a binary mixture of a solid colloidal phase, with volume fraction $\phi_{c} \equiv \phi$, and a liquid phase with volume fraction $\phi_{l}$. This composition can vary throughout space and time, but we will leave the dependence on position $\boldsymbol{r}$ and time $t$ of our fields implicit throughout to ease the notation. Volume conservation implies that $\phi_{l} = 1 - \phi$. The phases have a density $\rho_{c}$ and $\rho_{l}$, respectively. Imposing mass conservation for each phase, we obtain the following integral expressions
\begin{align} 
\label{eq:mliq} \int_{V} \mathrm{d}V \; \partial_{t} \left[ \rho_{l} (1-\phi) \right]  &= - \int_{\partial V} \mathrm{d}\boldsymbol{A} \cdot \left[ \rho_{l} (1-\phi) \boldsymbol{v}_{l} \right] , \\
\label{eq:msol} \int_{V} \mathrm{d}V \; \partial_{t} \left( \rho_{c} \phi \right) &= - \int_{\partial V} \mathrm{d}\boldsymbol{A} \cdot \left( \rho_{c} \phi \boldsymbol{v}_{c} \right) .
\end{align}
Here, $\boldsymbol{v}_{c}$ and $\boldsymbol{v}_{l}$ are the colloid and solvent velocity, respectively, `$\cdot$' indicates the inner product, and $\partial_{t}$ is the partial derivative with respect to time. Integrals on the left-hand side are over the volume of a stationary control volume $V$, while the right-hand side integrates the fluxes through that volume's surface $\partial V$. Since the control volume is in principle arbitrary, Eqs.~\eqref{eq:mliq} and~\eqref{eq:msol}, give rise to the following differential form
\begin{align} 
\label{volumebalance} &\partial_{t} \phi = -\boldsymbol{\nabla} \cdot  \left( \phi \boldsymbol{v}_{c} \right) , \\
\label{eq:ome} \boldsymbol{\nabla} \cdot &\left[\phi \boldsymbol{v}_{c} + (1-\phi) \boldsymbol{v}_{l} \right] = 0.
\end{align}
Equation~\eqref{volumebalance} implies that $\phi$ only changes \textit{via} flow-mediate flux through the boundary and we will use it to integrate $\phi$ in time. Equation~\eqref{eq:ome} defines a compressibility criterion on a volume-fraction weighted flow. 

\subsection{\label{sec:momentum}Momentum Balance Equation}

Next, we derive an expression for the velocity of the colloidal phase, relative to the liquid-phase velocity. First, consider the momentum balance equation for the entire binary mixture
\begin{align} 
\label{eq:mom} \frac{d}{dt} \int_{V} \mathrm{d}V \; \left[ \rho_{\mathrm{mix}} \ \boldsymbol{v}_{\mathrm{mix}} \right] &= \int_{V} \mathrm{d}V \; [\rho_{\mathrm{mix}} \ \boldsymbol{g} + \boldsymbol{\nabla} \cdot \boldsymbol{\sigma}_{\mathrm{mix}}] .
\end{align}
On the left-hand side we take the total time derivative of the momentum density, which follows from integrating the total density $\rho_{\mathrm{mix}}(\phi) = \rho_c \phi + \rho_{l} (1-\phi)$ together with the velocity of the mixture $\boldsymbol{v}_{\mathrm{mix}}$. By adding equations ~\eqref{eq:mliq} and~\eqref{eq:msol}, we can define a mixture flux $\boldsymbol{J}_m = \rho_m(\phi)\boldsymbol{v}_m,$ and subsequently write the mixture velocity as
\begin{align}
\label{eq:mix} \boldsymbol{v}_{\mathrm{mix}} &= \frac{\rho_{c} \phi \boldsymbol{v}_{c} + \rho_{l} (1-\phi) \boldsymbol{v}_{l} }{ \rho_{\mathrm{mix}}(\phi) } ,
\end{align}
which represents a weighted average by the densities of the two phases. On the right-hand side of Eq.~\eqref{eq:mom}, momentum is introduced into the system by the gravitational acceleration $\boldsymbol{g} = (0,0,-g)$ and internally redistributed \textit{via} (the divergence of) the mixture stress tensor $\boldsymbol{\sigma}_{\mathrm{mix}}$. We define the latter as a sum over a Newtonian term from the liquid solvent, and a non-Newtonian stress $\boldsymbol{\sigma}_c$, which is related to the presence of the solid phase in the mixture
\begin{align}
 \boldsymbol{\nabla} \cdot \boldsymbol{\sigma}_{\mathrm{mix}}  = -\boldsymbol{\nabla} p + \mu \ \boldsymbol{\nabla} \cdot \boldsymbol{\tau} + \boldsymbol{\nabla} \cdot \boldsymbol{\sigma}_c ,
\end{align}
Here, $p$ and $\mu$ are the liquid hydrostatic pressure and dynamic viscosity, respectively, and $\boldsymbol{\tau} = \boldsymbol{\nabla} \boldsymbol{v}_l + (\boldsymbol{\nabla} \boldsymbol{v}_l)^T$ represents the rate-of-strain tensor.

To make progress, we now consider the momentum balance equation for each phase separately. Introducing the momentum-exchange terms $\boldsymbol{\Sigma}_{ij}$ between phases $i,j \in \{l,c\}$, we arrive at
\begin{subequations}
\label{eq:mombal}
\begin{align}
\label{eq:momcol} \frac{d}{dt} \int_{V} \mathrm{d}V \; \left( \rho_c \phi \boldsymbol{v}_c \right) &= \int_{V} \mathrm{d}V \; (\rho_c \phi \boldsymbol{g} + \boldsymbol{\Sigma}_{cc} + \boldsymbol{\Sigma}_{cl}) \\
\label{eq:momliq}  \frac{d}{dt} \int_{V} \mathrm{d}V \; \left[ \rho_l (1-\phi) \boldsymbol{v}_l \right] &= \int_{V} \mathrm{d}V \; [\rho_l (1-\phi) \boldsymbol{g} + \boldsymbol{\Sigma}_{ll} + \boldsymbol{\Sigma}_{lc}].
\end{align}
\end{subequations}
Clearly, Newton's third law implies $\boldsymbol{\Sigma}_{cl} = - \boldsymbol{\Sigma}_{lc}$. Summing the two individual contributions above and subtracting equation~\eqref{eq:mom}, we obtain
\begin{align} 
 \boldsymbol{\nabla} \cdot \left[ - \mathbb{I} p + \mu \boldsymbol{\tau} + \boldsymbol{\sigma}_c + \rho_c \rho_l\frac{\phi(1-\phi)}{\rho_m(\phi)} \boldsymbol{\delta v}\boldsymbol{\delta v} \right] = \boldsymbol{\Sigma}_{ll} + \boldsymbol{\Sigma}_{cc}  ,
\end{align}
with $\mathbb{I}$ the identity operator, and $\boldsymbol{\delta v} \equiv \boldsymbol{v}_c - \boldsymbol{v}_l$ the net velocity of the colloids with respect to the background liquid flow. Evaluating the above equation in absence of colloidal phase, we can split the momentum exchange terms as
\begin{subequations} \label{eq:momex}
\begin{align} 
\label{eq:momexcol} \boldsymbol{\Sigma}_{cc} &= \boldsymbol{\nabla} \cdot \boldsymbol{\tilde{\sigma}}_c  \\
\label{eq:momexliq} \boldsymbol{\Sigma}_{ll} &=   \boldsymbol{\nabla}  \cdot (-\mathbb{I}p+ \mu \boldsymbol{\tau}),
\end{align}
\end{subequations}
where we have redefined the non-Newtonian stress 
\begin{align}
\boldsymbol{\tilde{\sigma}}_c = \boldsymbol{\sigma}_c + \frac{\rho_c \rho_l \phi(1-\phi)}{\rho_\mathrm{mix}} \boldsymbol{\delta v}\boldsymbol{\delta v} ,
\end{align}
to ease the notation.

Lastly, we chose to break up the cross terms into a hydrostatic contribution and a dynamical (friction) term
\begin{align}
\label{eq:momexcolliq} \boldsymbol{\Sigma}_{cl} &= -\phi \boldsymbol{\nabla} p + \boldsymbol{F}_{\mathrm{Darcy}} ,
\end{align} 
the latter of which we choose to model using (a variant of) Darcy's law~\cite{carman1939permeability,gray2004examination,carrillo2019darcy}. This law states that $\boldsymbol{F}_{\mathrm{Darcy}}$ is proportional to the relative velocities between the two phases, and it represents a coarse-grained contribution of all the microscale dissipative dynamic between the colloids and the solvent. Here, we write \begin{align}
\boldsymbol{F}_{\mathrm{Darcy}} = \frac{\mu}{\sigma^2} \frac{(1-\phi)(\boldsymbol{v}_{l} - \boldsymbol{v}_{c})}{K(\phi)}.
\end{align} 
The dimension-free scalar function $K(\phi)$ represents the porosity of the mixture, $\sigma = 2a$ the (mean) particle diameter, and $\mu$ the viscosity of the suspending fluid, which ensures the correct dimensionality. Note that we must have that $K$ diverges in the absence of colloids, such that the Darcy-like term disappears from the equation of motion. Conversely, $K$ must go to zero as $\phi$ approaches the maximum value allowed $\phi_m$ --- depending on the context this could be random loose packing, random close packing, or another dense arrested state. We will provide an explicit form for $K(\phi)$ in Section~\ref{sec:porosity}.

By defining a phase-related convective derivative $D^i_t= \partial_t +\boldsymbol{v}_i \cdot\boldsymbol{\nabla}$, and combining equations~\eqref{eq:mombal} with~\eqref{eq:momex} and~\eqref{eq:momexcolliq}, we can eliminate the pressure $p$ from the system. This allows us to arrive at our final relation
\begin{align} 
\label{eq:navstok}
D^c_t[\rho_c \boldsymbol{v}_c] - D^l_t[\rho_l  \boldsymbol{v}_l]  + \frac{ \mu \boldsymbol{\delta v}} {\sigma^2 \phi K} 
= \Delta\rho \boldsymbol{g} + \frac{\boldsymbol{\nabla} \cdot \boldsymbol{\tilde{\sigma}}_c}{\phi} - \frac{\mu \boldsymbol{\nabla} \cdot \boldsymbol{\tau}}{1-\phi} .
\end{align}

\subsection{\label{sec:quasihydrostatic} Quasi-Hydrostatic Limit}

We perform a dimensional analysis on Eq.~\eqref{eq:navstok} and conclude that there it is useful to introduce a characteristic velocity $v_g$ and length $l_R$. The former represents the bulk settling velocity of a single colloid due to buoyant forces $v_g = \Delta\rho g \sigma^2/ \mu$. The latter is the length scale associated with smallest distance that the coarse-grained theory can resolve. That is, $l_R^3$ is the Representative Elementary Volume (REV) ~\cite{carrillo2019darcy}, which is the smallest volume containing sufficient particles to define a local field $\phi$. This parameter is system dependent and can be used to demarcate the scale transition from micro- to macro-scale descriptions in porous materials~\cite{gray2004examination}.

From our dimensional analysis, we conclude that Eq.~\eqref{eq:navstok} can be rewritten as
\begin{align} 
\lim_{\mathrm{Re} \rightarrow 0} \ \frac{ \boldsymbol{\delta v}^*} {K(\phi )} 
=  \ -\phi \boldsymbol{\hat{z}} + \alpha^2 \left[ \boldsymbol{\nabla}^* \cdot \boldsymbol{{\sigma^*}}_c - \frac{ \phi \boldsymbol{\nabla}^* \cdot \boldsymbol{\tau}^*}{1-\phi} \right ].
\end{align}
Here, we have introduced the dimensionfree constants $\alpha = \sigma / l_R$ --- the ratio of the particle diameter to the REV size --- and $\mathrm{Re} = \Delta \rho v_g l_R / \mu$ --- the Reynolds number. Quantities denoted with the $\ast$ superscript are expressed in natural units. The limit $\mathrm{Re} \rightarrow 0$ naturally emerges, since in real systems this limit holds for the small particles suspended in a viscous medium.

We can further simplify the expression by considering the limit as $\alpha \rightarrow 0$. This represents highly correlated systems, where the amount of particles needed to be described at a coarse-grained level is extremely large. However, caution is advised in following this approach. For finite-size systems of typical length $L$, $\alpha$ is naturally bounded from below by the ratio $\sigma/L$. Lower values would make the coarse-grained description meaningless. Moreover, the non-Newtonian contribution to the stress $\sigma_c$ may be singular (divergent), particularly at the maximum packing fraction $\phi_{m}$. This, for example, prevents the non-physical overlap of hard colloids. This makes taking the limit $\alpha \rightarrow 0$ poorly defined in some cases.

The limit wherein $\alpha$ and $\mathrm{Re}$ both tend to zero, is commonly referred to as the quasi-hydrostatic regime. Assuming small values of $\alpha$ and $\mathrm{Re}$, we can express
\begin{align} 
\label{eq:qseqn}
\boldsymbol{\delta v} 
\approx K(\phi ) (-v_g \phi \boldsymbol{\hat{z}} + \sigma^2 \mu^{-1}     \boldsymbol{\nabla} \cdot \boldsymbol{{\sigma}}_c).
\end{align}
At this point, we need to specify the forms of $K(\phi)$ and $\boldsymbol{\sigma}_c$ in order to solve the problem. This also requires us to combine equation~\eqref{eq:qseqn} with~\eqref{eq:ome} to eliminate the dependence on the liquid velocity, as we will show in Section~\ref{sec:res}.    

\subsection{\label{sec:porosity}Mean Porosity}

The distribution of pore cross-sections within the colloidal phase is intricately linked to the distribution of particle surface-to-surface distances. This relationship is typically hard to predict~\cite{carman1939permeability,heijs1995numerical,xu2008developing,ozgumus2014determination}, as it is influenced by factors such as system temperature, details of interparticle interactions~\cite{ruiz2020tuning}, and potentially, the system's preparation/rheological history~\cite{koumakis2015tuning,gibaud2020rheoacoustic,Garbin-gel,torre2023structuring}. For simplicity, and as a first-order approximation, we opt to characterize the mean pore cross-section using the following ansatz
\begin{align} 
\label{eq:K}
K(\phi ) = \frac{(1-{\phi} / {\phi_{m}})^3}{k^2_0 \phi} ,
\end{align}
in this work. This expression adequately reproduces the findings reported in~\cite{torquato1995mean}, where the authors estimate the nearest-neighbor distance in a homogeneous configuration of hard spheres as a function of the volume fraction $\phi$. Thus, we model the mean pore cross-section simply as the mean surface-to-surface distance between spherical particles. The constant parameter $k_0 = 3\sqrt{2}$ is determined by computing the sedimentation velocity of a single sphere in a viscous fluid from equation~\eqref{eq:qseqn}.
    
\subsection{\label{sec:viscosity}Dilatational Viscosity}

We choose to represent the non-Newtonian stress as a simple diagonal operator, which depends on the local $\phi$ and the amount of compression in the colloid field  
\begin{align} 
\label{eq:sigma_c} \boldsymbol{\sigma}_c = \lambda(\phi) \mathbb{I} \left( \boldsymbol{\nabla} \cdot \boldsymbol{v}_c \right).
\end{align}
Here, $\lambda$ denotes the dilatational viscosity, which physically represents the irreversible resistance to compression or expansion within a fluid. At the microscopic level, it arises from the finite time needed for energy injected into the system to disperse among the various degrees of freedom of colloidal motion~\cite{landau1959fluid}. Note that Eq.~\eqref{eq:sigma_c} does not account for shear stress response of the colloidal phase. However, in the present study we limit our focus to one-dimensional (1D) systems, where the shear component of the stress is not resolved.

We model the dilatational viscosity as the product of a local energy density $\epsilon(\phi)$ and relaxation time $T_{\mathrm{r}}(\phi)$. The first can be expressed as the product of particle density $n_c = N_c/V$ and the average number of bonds per particle $n_b(\phi)$. To estimate the latter, we compute the ratio of incoming $j_{\mathrm{in}}=D_0 / \delta r^2(\phi)$ and outgoing $j_{\mathrm{out}}=D_0 / \Gamma^2$ fluxes of particles for a single central colloid. Here, $D_0 = (k_{\mathrm{B}} T) / (3 \pi\mu \sigma)$ represents the diffusion coefficient of an isolated colloid in an unbounded fluid, and $\Gamma \gtrsim \sigma$ denotes the range of the attractive interactions, as measured from the colloid centers. We retain only the fraction of particles that do not escape from the potential well of the central colloid, resulting in
\begin{align} 
\label{eq:epsilon}
\epsilon(\phi) =  \epsilon_0  n_c n_b(\phi) = \epsilon_0 \frac{3 z_{\mathrm{max}} }{\pi \sigma^3} \frac{ \phi  \Gamma^2}{\delta r^2(\phi)} (1-e^{-\epsilon_0 / k_B T }). 
\end{align}
Here, $z_{\mathrm{max}}$ represents the maximum number of bonds a particle can form due to geometric constraints, and $\epsilon_0$ is the energy of a single bond. Additionally, and similarly to what we assumed for the mean porosity, $\delta r = \sigma [1+ (1-\phi/\phi_m)^{3/2}\phi^{-1/2}/10 ]$ denotes the mean distance to the closest neighboring particle in a spatially homogeneous distribution of colloids, approximately reproducing Torquato's original result~\cite{torquato1995mean}.

Next, we calculate the relaxation time of the colloidal phase. To obtain an estimate, we use an effective diffusion coefficient $D(\phi)=D_0 f(\phi)$ and expand it to first order around $\phi=\phi_{m}$ where diffusion is suppressed by crowding, thus yielding $D |_{\phi=\phi_m} = 0$:
\begin{align}
\label{eq:relaxtime}
T_{\mathrm{r}}(\phi) =\frac{l^2_R}{D_0 (1-\phi/\phi_m)}. 
\end{align}
Now we are in a position to combine Eqs.~\eqref{eq:epsilon} and~\eqref{eq:relaxtime} to obtain the dilatational viscosity
\begin{align} 
\lambda(\phi) =\mu \frac{9 \ \alpha^{-2} \ z_{\mathrm{max}} \ U(1-e^{-U})}{[\phi^{1/2}+\frac{1}{10} (1-\phi/\phi_m)^{3/2}]^2} \ \frac{ \phi^2 }{(1-\phi/\phi_m)}, 
\end{align}
where we introduce the dimension-free potential strength $U=\epsilon_0 / k_B T$, and assume short-range interactions ($\Gamma \approx \sigma$). As expected, the stress exhibits a singularity at the maximum packing fraction $\phi_m$. This implies it can serve as a means to dissipate kinetic energy at a rate large enough to suppress applied stresses. In turn, this would enable the (dense) gel to support its own weight.

\subsection{\label{sec:helmholtz}Helmholtz Decomposition}

The above framework is limited to 1D systems. When extending our analysis to higher dimensions, an additional relationship involving the spatial derivative of the colloid velocity becomes necessary to close the system. One possible approach is to perform a Helmholtz decomposition~\cite{ribeiro2016helmholtz, klaseboer2019helmholtz, glotzl2023helmholtz} on the mixture velocity field. That is, we can express it as
\begin{align} 
\boldsymbol{v}_{\mathrm{mix}} = \boldsymbol{\omega} + \boldsymbol{\zeta}, 
\end{align}
where the vector field is divided into a divergence-free component, $\boldsymbol{\omega}$, and a curl-free component, $\boldsymbol{\zeta}$. A natural choice for the former is derived from equation~\eqref{eq:ome}, namely $\boldsymbol{\omega} = \phi \boldsymbol{v}_{c} + (1-\phi) \boldsymbol{v}_{l}$. Consequently, we can explicitly compute $\boldsymbol{\zeta} = \boldsymbol{v}_{\mathrm{mix}} - \boldsymbol{\omega}$, and obtain the following constraint
\begin{align} 
\boldsymbol{\nabla} \times &\left [ \frac{\Delta \rho}{\rho_{\mathrm{mix}}} \phi (1-\phi) \boldsymbol{\delta v} \right ] = 0.
\end{align}
It is worth noting that this decomposition is not always unique and may not be applicable in all scenarios~\cite{glotzl2023helmholtz}. This is because the smoothness of the mixture velocity field $\boldsymbol{v}_{\mathrm{mix}}$ is a prerequisite for employing this technique.

\section{\label{sec:res}Results}

In this section, we explore solutions of our model specifically tailored to 1D systems. We start our analysis by deriving the exact solution for the initial ($t = 0)$ colloid velocity field $ \boldsymbol{v}_c$, when there is a homogeneous density profile. This already provides key insights into the origin of the gel's strength, and how it is able to support its own weight (for a finite time). Subsequently, we consider the full time evolution numerically. This clearly demonstrates the gel's resistance to gravitational collapse and reproduces gel settling regimes qualitatively.

\subsection{\label{sec:1Dan}Homogeneous Colloid Distribution}

Consider an homogeneous isolated system with net volume $HL^2$ and height $H \ll L$, as well as initial colloid volume fraction $\phi(z) |_{t=0} = \phi_0$. Due to the underlying symmetry, the velocity field can be expressed as a function of the $z$-coordinate and time only $\boldsymbol{v}_c = v_c(z,t) \boldsymbol{\hat{z}}$. Combining equation~\eqref{eq:ome} with~\eqref{eq:qseqn}, and defining a characteristic length
\begin{align}
\zeta= \sigma \sqrt{K(\phi_0)\lambda(\phi_0)(1-\phi_0) / \mu} ,
\end{align} we obtain the following equation for the initial colloid velocity field
\begin{align} \label{eq:homog_1d}
 \partial_z {v}_c |_{t=0} \ = \ \zeta^2 \ \partial^3_z {v}_c |_{t=0} .
\end{align}
Equation~\eqref{eq:homog_1d}, when extended to arbitrary dimensions, reveals its nature as a Poisson equation describing the stress $\boldsymbol{\sigma}_c$, with a source term directly linked to the gel's compression, expressed as $\phi \boldsymbol{\nabla} \cdot \boldsymbol{v_c}$.

Using mass conservation, we integrate Eq.~\eqref{volumebalance} in space to obtain the boundary conditions
\begin{align} 
v_c|^{z=0,H}_{t=0}&=0 , \\
\left. \partial^2_z v_c  \right |^{z=0,H}_{t=0} &= \frac{v_g \phi_0}{\lambda(\phi_0)}.
\end{align}
We can also define a characteristic colloid velocity $V_g = v_g \phi_0 (1-\phi_0)K(\phi_0)$, representing the colloidal flow in absence of non-Newtonian stress. The presence of a characteristic length scale $\zeta$ and velocity $V_{g}$ allows us to define a characteristic time $\tau_d = \zeta / V_g$. In the next section, we will show that the latter can be used to obtain an accurate estimation of the gel collapse delay time.

Finally, combining the above results, the initial colloid velocity can be written as 
\begin{align} 
\label{eq:v_c_init} {v}_c(z)|_{t=0} = V_g  \left [ \frac{\cosh{\frac{H/2 - z}{\zeta}}}{\cosh{\frac{H}{2\zeta}}} - 1 \right ].
\end{align}
Here, we recognize $\zeta$ as a stress screening length for the gel. In the limit of $\zeta / H \gg 1$, a strong resistance to compression and expansion is extended to the entire sample, halting the sedimentation process. Note that, a similar concept of stress screening length was already introduced by Evans and Starrs~\cite{evans2002emergence}.

\subsection{\label{sec:1Dnum}Time-Evolution}

For our numerical analysis, we set the height of our sample $H=5 \cdot 10^{-2} \mathrm{m}$, and the
bulk settling velocity of a single colloid $v_g = 10^{-5} \mathrm{m} \mathrm{s}^{-1}$, reproducing commonly used experimental setups~\cite{allain2001systematic,  allain2003rapid, weitz2005gravitational,zaccarelliHarich2016}. We also set $l_R \approx 1.3 \ 10^{-4}$m, and incrementally discretize height using a spacing $\delta z \in \{ 10^{-3}, 5 \times 10^{-4}, 2.5 \times 10^{-4} \}H$, where the smaller values are used for more numerically challenging systems. For the initial volume fraction, we choose $\phi_0 \in [0.01, 0.55]$, which spans the range of experimentally considered densities~\cite{zaccarelliHarich2016,darras2022}, and set $\phi_m=0.7$. Finally, we take the (reduced by $k_{\mathrm{B}}T$) potential strength in the range $U \in [5, 50]$. Here, $U \approx 5$ is a typical lower bound for colloidal gelation~\cite{zaccarelliHarich2016}, while $U \approx 50$ represents a very strong gel at intermediate to high $\phi_{0}$.

We numerically solve equations~\eqref{eq:ome} and~\eqref{volumebalance} for a 1D system with these choices, to obtain the colloidal density profile as a function of time; see Appendix~\ref{sec:appen} for more information. For these solutions, we define the height of the gel interface $h(t)$ --- between the colloid-rich and colloid-poor part of the sample --- as the tallest point in the system where the local colloid volume fraction $\phi(z)$ exceeds $\phi_0/2$.

\begin{figure}[!htb]
\centering
\includegraphics[width=85mm]{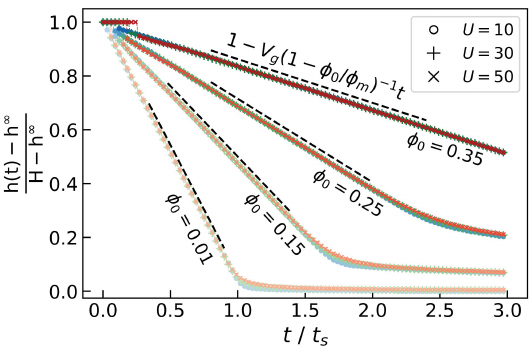}
\caption{\label{fig:h_t}Time evolution of the interface height between the settling gel and supernatant region. Lighter colors denote less dense configurations, while darker colors represent denser ones, as labelled, with the symbols indicating various bond strengths, see the legend. The interface height is adjusted by its long-term limit $h^{\infty} = H \phi_0 / \phi_m$ (complete separation) and scaled to fit within the range of $[0,1]$. Dashed black lines are guides to the eye to indicate a linear trend with slope $(1-\phi_0/\phi_m)^{-1} V_g(\phi_0) $. Time is expressed in units of $t_s = H / v_g$, representing the time required for a single colloid to sediment the height of the sample volume $H$.}
\end{figure}

Figure~\ref{fig:h_t} shows $h(t)$ for several choices of $\phi_0$ and $U$. We observe three regimes: an initial delay, a linear `collapse', and an exponential compaction. The former is hard to distinguish for low $\phi_0$, while the latter cannot be observed for high $\phi_0$, due to the slow settling dynamics for such $\phi_0$. Examining the delay more closely, we find that it is significantly influenced by both $\phi_0$ and $U$, which reflects the monotonic increase of the dilatational viscosity $\lambda$ with these parameters. An analytical expression for the delay time will be provided later in this section, in Eq.~\eqref{eq:theory_delayt}. In linear regime, the interface settles at a constant velocity $(1-\phi_0/\phi_m)^{-1}V_g$, determined by the porous structure of the gel medium rather than its resistance to deformations. The onset of the third regime occurs when the interface reaches the dense region formed at the bottom of the sample, with the settling of the (largest part of the) gel bulk. In this regime, sedimentation speed is predominantly governed by colloid-solvent hydrodynamic drag, with minimal deviations observed among systems with different potential strengths $U$.

\begin{figure}[!htb]
\centering
\includegraphics[width=80mm]{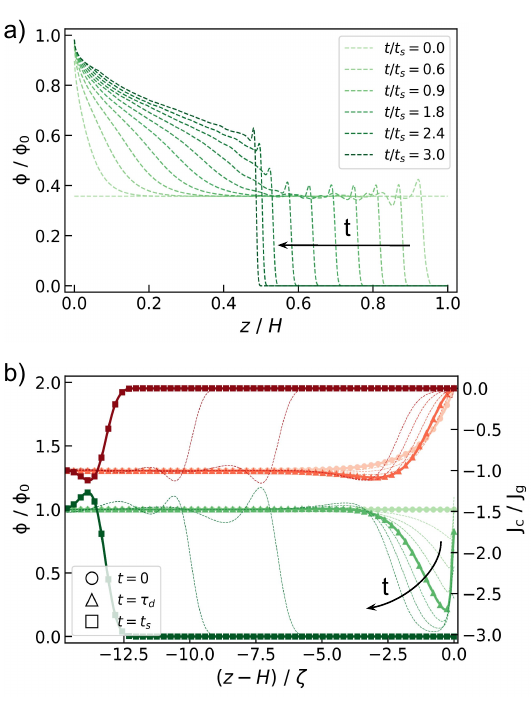}
\caption{\label{fig:dens_prof}Density profiles depict the time-evolution of a sedimenting gel with an initial volume fraction $\phi_0=0.25$ and dimension-free potential strength $U = 30$. Local volume fractions $\phi(z)$ are normalized by their initial value $\phi_0$, with colors indicating temporal evolution from light to dark. The evolution direction is indicated by the black arrow. (a)~Colloid density sampled uniformly in time as a function of the reduced height $z/H$, with the time interval between each curve given by $0.3 t_s$. (b) The flux of colloids $J_c = v_c \phi$ (red), normalized by the bare gravitational flux $J_g = V_g \phi_0$, alongside normalized local volume fraction (green), plotted against height shifted by the system size $H$, and expressed in units of the gel stress screening length $\zeta$. Thick lines and symbols denote the system at three representative times: the initial configuration (circles), the creation of debris at the top of the sample (triangles), and during linear collapse (squares). The thin lines give an impression of the behavior at intermediate times $\{0.02,0.04,0.07,0.09,0.13,0.15,0.5, 0.75 \}$ $t_s$.}
\end{figure}

Next, we connect the (early time) behavior of the interface to the gel's internal structure. Figure~\ref{fig:dens_prof}a shows the time evolution of density profiles for a sedimenting gel with $\phi_0=0.25$ and $U = 30$. In the linear settling regime, the bulk of the gel sediments almost unperturbed. That is, there is a large section of the gel that remains at $\phi_0$. We also see a small (physical) peak at the upper end of this flat range, which represents the accumulation of `debris' on top of the gel. We will see that this originates from the early stages settling when we turn to Fig.~\ref{fig:dens_prof}b.

Turning our attention to the bottom of the sample, we see that here the gel begins to compact immediately. That is, for our parameter choices, the gel cannot fully support its own weight, we will return to this in Section~\ref{sec:disc}, when we discuss the connection to experiment. The  transition from linear collapse to exponential settling occurs around $t \approx 2.1 t_{s}$, and can be seen to coincide with the dense region at the bottom meeting with the interface between the gel and supernatant.

In Fig.~\ref{fig:dens_prof}b, early-time colloidal flux $J_c (z) = \phi v_c (z)$ and $\phi(z)$ are depicted. As the bulk of the system sediments, it stretches the top part and compresses the bottom part of the gel --- referencing Eq.~\eqref{eq:v_c_init} the top and bottom can be identified to be within $\zeta$ of either boundary. The top of the gel tries to resist this stretch until the thinnest part connecting it with the bulk breaks. Unsurprisingly, this is roughly when the bulk has settled a distance $\zeta$. This leaves the top part detached from the rest of the gel. Note that here we did not explicitly put in an attraction between the top of the sample and the gel, the apparent attachment to the top of the gel is caused by the no-flux boundary conditions imposed on the colloid velocity field $v_c$. These are necessary to ensure the system is isolated.

Throughout the evolution of the system, the part that remains at the top decreases in density and creates debris. These will sediment at a faster rate than the denser bulk gel due to the weaker hydrodynamic dissipation this debris experiences. Ultimately, the debris deposits itself on top of the bulk as it continues to sediment and causes a local density peak. As this denser material resists settling more, a local density minimum eventually forms in front of it. Revisiting Fig.~\ref{fig:h_t}, we understand that the delayed collapse can be attributed to the top part of the gel resisting collapse for a finite amount of time. Considering $J_c(z)$ in Fig.~\ref{fig:dens_prof}b, the top layer detaches from the gel bulk, the flux of colloids transforming from a monotonically increasing function to a function with local minima and maxima. The former indicates that the gel is swelling everywhere near the top boundary in a cohesive way. The latter signals that a part of the gel has detached from the bulk and is accumulating on top of the homogeneously collapsing bulk.

\begin{figure}[!htb]
\centering
\includegraphics[width=85mm]{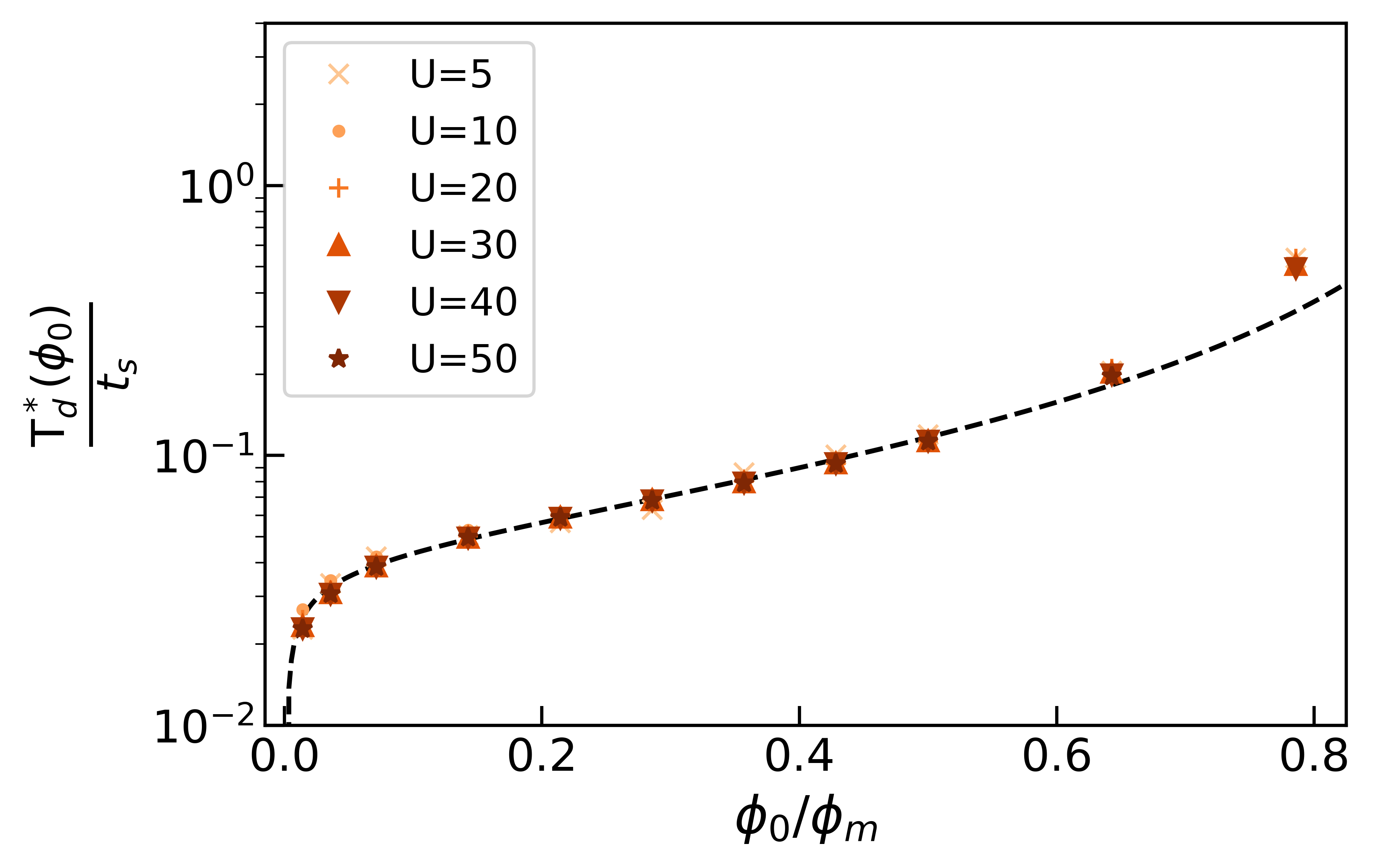}
\caption{\label{fig:delay_time}Delay time for gravitational gel collapse $T_d^{\ast}$, in units of $t_s$, as a function of initial volume fraction $\phi_0$, and scaled by $c(U)=\sqrt{U(1-e^{-U})}$. Numerical results for different values of the potential strength $U$ are indicated using symbols, see legend. The dashed line represents the theoretical prediction~\eqref{eq:theory_delayt}, scaled in the same way as the data points.}
\end{figure}

Finally, in Figure~\ref{fig:delay_time} we plot the delay time $T_d$ obtained from numerical solutions --- defined as the earliest time for which $h(t)<H$ --- as a function of initial volume fraction $\phi_0$. The delay time is scaled by $c(U)=\sqrt{U(1-e^{-U})}$ to ensure data collapse. This scaling implies that the main contribution of stronger interactions is an increase in the delay time, while the underlying physics is not modified, as confirmed by the data collapse in Fig.~\ref{fig:h_t}. We compare the numerical results with our theoretical prediction 
\begin{align} 
\label{eq:theory_delayt}
T_d = a_0 \left ( 1-\frac{\phi_0}{\phi_m} \right ) \tau_d,
\end{align}
where $a_0 \approx 5/2$ is a fit parameter likely resulting from the way in which we define the gel interface in our system. We obtained this by computing the time required for an interface moving at a constant velocity $V_g(1-\phi_0 / \phi_m)^{-1}$ to cover a distance $\zeta$. This result matches our numerical data extremely well across the range of $U$ and $\phi_0$ considered. Turning to our simple description of the early-stage settling, we conclude that the bulk of the gel must settle a distance of at least $\zeta$ to free itself from the non-Newtonian responses of the boundary. Once it does, it is able to settle at a constant velocity $V_g$. We will provide further elaboration on this point, when we provide additional context to our research in the next section.

\section{\label{sec:disc}Discussion}

We have put forward a theoretical description of a colloidal gel, where we have moved away from previous modeling~\cite{buscall1987consolidation, evans2002emergence,weitz2005gravitational,buzzaccaro2012equilibrium,darras2022} by introducing a dilatational viscosity for the solid phase. This imparts resistance to compression and expansion, which strongly affects the gel's settling behavior under gravity. In this section, we discuss our results in the context of experimental studies~\cite{allain2003rapid,weitz2005gravitational,zaccarelliHarich2016,zhou2018onset,darras2022,john2024early} and other computational~\cite{varga2018modelling,padmanabhan2018gravitational,de2023hydrodynamic}, and theoretical work~\cite{evans2002emergence,allain2003rapid,weitz2005gravitational,darras2022}.

\subsection{\label{sec:layer}Layering and Length Scales}

Our $t = 0$ analytic expression, Eq.~\eqref{eq:v_c_init}, provides insight into our numerical observations, which we consider valid on the basis of the correspondence in Fig.~\ref{fig:delay_time}. A homogeneous initial sample can be roughly divided into three regions: a top and bottom layer, both with a thickness of $\zeta$ --- the gel stress screening length --- and a bulk containing the rest of the sample. As this homogeneous gel begins to sediment, the top and bottom layers resist stretching and compression, respectively, while the bulk settles with a uniform velocity $V_g$.

In our calculations, the bottom part typically lacks the strength to withstand compression, leading to the immediate creation of a bottom dense layer that grows without delay (see Figure~\ref{fig:dens_prof}a). This indicates that the gel stress screening length, $\zeta$, is typically too small to propagate the stress response from the bottom of the sample to a height sufficient for forming a resistant layer thick enough to prevent the gel bulk from collapsing. A bottom layer, that can be clearly distinguished from the bulk gel for an extended period of time, is also observed in certain experimental configurations~\cite{zaccarelliHarich2016}. These are usually associated with gels having strong inter-particle interactions, suggesting that their size depends on the strength of the attractions, which aligns with our findings for $\zeta$.

Notably, a similar characteristic stress screening length is also reported in the work of Evans and Starrs~\cite{evans2002emergence}, where a comparable theoretical framework is applied. Their study demonstrated that the characteristic velocity of the collapse could be influenced by the size of this screening length, particularly when the sample size is comparable to it. This suggests that gels with extremely long delay times~\cite{zaccarelliHarich2016,zhou2018onset,darras2022} may fall into this category. In such cases, the velocity of the colloids is exponentially suppressed throughout the entire sample, resulting in minimal or no dynamics, and consequently, a significant delay in the onset of collapse.

\subsection{\label{sec:rupture}Settling and Rupture}

The initial settling causes an upward solvent flow (backflow), with the velocity primarily dictated by the porous structure of the networked colloids. We believe this to be realistic for unruptured gels in experimental settings~\cite{darras2022}, as well as for colloids with small sizes~\cite{allain2003rapid, weitz2005gravitational}. In these systems, the interface velocity follows a power law that is different from the one in our model, due to the specific choice for porosity $K$. The scaling predicted by our theory follows from the assumption that pores are locally homogeneously distributed within a representative elementary volume (REV), thus recovering the expression for a single isolated sphere sedimenting in an unbounded Newtonian solvent. A more accurate porosity for fractal-like aggregates should also depend on the fractal dimensions of these aggregates~\cite{allain1995aggregation, allain2003rapid, weitz2005gravitational}. However, estimating such a quantity is far from trivial, as it generally depends on the volume fraction, the strength of the interactions, and the shape of the particles~\cite{allain2003rapid}, making it space-dependent in our description. Therefore, we choose to use a simpler and well-defined expression as defined in~\eqref{eq:K}, which effectively covers the entire range of volume fractions and connects different regimes with a unified qualitative description.

When the center of mass of the bulk displaces by a height of $\zeta$, the top part of the gel is maximally extended. This triggers a detachment of the bulk gel, after which the remaining top part effectively becomes debris. These debris subsequently deposit on the gel interface at the top of the bulk area as it continues to settle. This accumulation of debris at the top of the gel interface has been observed in experiments~\cite{zaccarelliHarich2016,zhou2018onset}. In most experiments, there is a curved meniscus to the sample and the mechanism by which the debris forms is presumably not identical. However, the inverted-cuvette setup of Ref.~\cite{zhou2018onset} comes closest to our 1D theoretical analysis. Here, a denser layer on top of a uniformly settling gel was observed, though the formation of a thinner region before rupture could not be discerned. It may be that $\zeta$ is very different in these experiments or that the boundary conditions we considered are not commensurate with the ones of Ref.~\cite{zhou2018onset}.

Once the interface has detached from the top of the sample, it continues to sediment with constant velocity until it reaches the dense bottom layer. The interface subsequently slows down exponentially as the sample compacts and $h$ approaches its asymptotic limit $h^{\infty}$. The presence of such a compaction regime is also consistent with experimental observations and previous modeling~\cite{allain2003rapid, weitz2005gravitational, zaccarelliHarich2016,razali2017effects, darras2022,de2023hydrodynamic}. However, we do not believe that the constant velocity regime is representative of the ruptured regime that is encountered in experiments~\cite{allain2003rapid,zaccarelliHarich2016,zhou2018onset}. That is, the transition from delayed to linear collapse is widely accepted to occur when cracks~\cite{allain2003rapid}, or streamers~\cite{zaccarelliHarich2016,zhou2018onset}, emerge in a previously homogeneous bulk gel. They could either give preferential pathways for fluid flow or yield the gel network in its entirety~\cite{zaccarelliHarich2016}, due to erosion by fluid backflow~\cite{varga2018modelling,zhou2018onset, john2024early}. Our model gel remains homogeneous and its bulk stress is relatively unaffected before compaction occurs.

\subsection{\label{sec:deltimes}Extending the Delay Time}

As we do not have cracks or rupture, we might expect our delay times to be longer than in experiment, as also suggested by the inverted-cuvette experiment in Ref.~\cite{zhou2018onset}. However, $T_{d}$ is consistently shorter than reported in real systems~\cite{zaccarelliHarich2016,zhou2018onset,darras2022}. Crucial to understanding the origin of this difference is the form of our dilatational viscosity $\lambda(\phi,U)$. Scaling this viscosity by the function $c(U)$, as shown in Fig.~\ref{fig:delay_time}, preserves its dependency on $\phi_0$. This is evidenced by the data collapse in this figure. Suppose we had a different form of $c(U)$, we would expect the collapsed curve to still hold, but the real delay time change significantly. In proposing a form of $\lambda$, we ignored the influence of attractive interactions on the system's relaxation time. These should strongly suppress colloidal diffusion, thereby increasing drastically $c(U)$ and $T_{d}$. Incorporating a more detailed expression for $\lambda$ capable of capturing these effects should yield delay times more closely aligned with experimental values, and could be considered in follow-up studies.

Lastly, we note that the computational study by Padmanabhan and Zia~\cite{padmanabhan2018gravitational} predicted both a delayed and a collapse without accounting for hydrodynamic interactions. They attribute their observed delay to gravity-enhanced coarsening driven by negative osmotic compressibility, a mechanism not captured in our modeling. However, their predicted delay times are significantly shorter than those observed in experiments, and the range of gravitational P{\'e}clet numbers $\mathrm{Pe}_{g}$ they investigated is limited to small values, whereas not all experimental systems fall within this range~\cite{zaccarelliHarich2016, zhou2018onset}. In contrast, our model assumes that, in the absence of gravity, there would be no osmotic contribution to the stress affecting local density. We consider the dynamics to be dominated by gravity and the viscoelastic response of the gel network, mediated by hydrodynamics, making the P{\'e}clet number effectively infinitely large, as reflected by the missing diffusive term in Eq.~\eqref{volumebalance}.

\subsection{\label{sec:nD}Going beyond One Dimension}

The experiments carried out in the Poon lab~\cite{meeker1998low,starrs2002collapse,zaccarelliHarich2016,zhou2018onset} suggest that a 1D description may lack necessary features to accurately replicate real gel behavior. Missing features include the curvature of the suspension-air meniscus, which represents the top boundary of the system, the interaction between the gel and the meniscus, as well as between the gel and the other, solid confining boundaries~\cite{allain2001systematic,evans2002emergence}. That is, tangential slip of colloids along a curved meniscus could lead to local weakening of the structure at the top of the sample, and debris accumulation at specific points, which subsequently yield the gel and trigger the onset of rapid collapse~\cite{zhou2018onset}.

Additionally, the influence of shear-stress response in the gel does not play a role in our 1D calculations. This response is expected to be relevant for the same reasons that we mentioned, when we introduced the dilatational stress. We expect that shear-stress response could potentially produce more resistant gels, as any variation in the flow of colloids in the plane normal to its direction, would generate a response. These variations can be caused by microscale factors like density fluctuations coming from trapped solvent droplets. They can also stem from the applied boundary conditions,~\textit{e.g.}, the presence of a liquid-air meniscus~\cite{zaccarelliHarich2016,zhou2018onset} or an overall incline of the cuvette~\cite{allain2001systematic}.

Finally, our model assumes an instantaneous stress response for the gel network. However, in reality, the gel network undergoes erosion, which changes the morphology of the network and weakens its stress response over time~\cite{varga2018modelling, de2023hydrodynamic}. This erosion also alters the pore distribution, affecting the porosity $K$. That is, a gel that is initially much stronger than what we considered in this work, will weaken over time, leading to a more pronounced and violent transition from delayed to linear collapse. Future work will focus on the implementation of these `weakening' features.

\section{\label{sec:conclusion}Conclusions and Outlook}

Summarizing, we have developed a theoretical model, by which we can study the response of colloidal gels to gravitational stress. Our approach treats the gel as a viscoelastic medium, incorporating a dilatational viscosity for the colloidal phase that varies with local density. Herein, we depart from previous modeling~\cite{buscall1987consolidation, evans2002emergence,weitz2005gravitational,buzzaccaro2012equilibrium,darras2022}, which allows us to qualitatively capture several experimentally observed features of colloidal gel sedimentation. The gel elasticity is complemented by Darcy's law, by which we represent flow as a local drag force acting on the solid phase. That is,  we work at a level that coarse-grains out the complex micro-scale hydrodynamic flows and interactions.

We computed the full dynamics of the system numerically in an effective 1D geometry,~\textit{i.e.}, only the sample height is relevant. This reveals that our model predicts three distinct regimes: a delay, followed by linear settling, and a final exponential compacting. All of these elements have been reported in experimental systems~\cite{zaccarelliHarich2016, zhou2018onset, darras2022}. Surprising, solving analytically for the initial ($t = 0$) behavior is sufficient to understand the origin of the salient features of the full collapse dynamics in our model. This analysis reveals that there is a natural length scale $\zeta$ in the system, representing a gel stress screening length. Thus, a sample can be divided into three regions, when $\zeta < H$: a top and bottom layer and a middle `bulk' part of the gel. The latter remains relatively unaffected during the initial part of the settling, again mirroring experiments~\cite{zaccarelliHarich2016,zhou2018onset}. However, when the former is fully stretched, it triggers the onset of the linear settling regime and the subsequent formation of a dense layer of debris on top of the settling gel. This is reminiscent of some experimental observations that were made on depletion-based  colloidal gels~\cite{zhou2018onset}.

However, the present modeling does not capture all settling behaviors observed in experiment. In part, this can be attributed to the approximations we made in determining the resistance of the gel to flow. This makes the delay time much shorter than is typically observed in experiment. However, we provide a means to amend this discrepancy through suitable modification of the dissipative term. In part, the choice of modeling the dynamics in 1D proved useful to gain basic insight, but it fails to capture some of the more essential features of the experiments. For example, we do not account for the the curvature of a liquid-air meniscus, which appears to control much of the onset of rapid collapse~\cite{zaccarelliHarich2016,zhou2018onset}.

Addressing these gaps in our present modeling could bring our results in closer agreement with the experiments. Future work will therefore extend the analysis to higher dimensions and incorporate erosion mechanisms. The present study provides a solid foundation for this and may be readily built upon in other directions.

\section*{Acknowledgements}

The authors acknowledge NWO for funding through OCENW.KLEIN.354. We are grateful to Dr. Alexis Darras, Myrthe van Leeuwen and Martijn van Schaik for useful discussion. An open data package containing the means to reproduce the results of the simulations is available at: [DOI]

\section*{Author Contributions}

Author contributions: Conceptualization, J.d.G. \& K.W.T.; Methodology, K.W.T.; Numerical calculations, K.W.T., Validation, K.W.T.; Investigation,
K.W.T.; Writing --- Original Draft, K.W.T.; Writing --- Review \& Editing, J.d.G.; Funding Acquisition, J.d.G.;
Resources, J.d.G.; Supervision, J.d.G.

\bibliographystyle{aip}
\bibliography{reference}

\begin{thebibliography}{10}

\bibitem{lekkerkerker1992poon}
H.~N. Lekkerkerker, W.-K. Poon, P.~N. Pusey, A.~Stroobants, and P.~. Warren,
\newblock Europhysics Letters {\bf 20}, 559 (1992).

\bibitem{poon2002physics}
W.~Poon,
\newblock J. Phys. Cond. Mat. {\bf 14}, R859 (2002).

\bibitem{bergenholtz2003gelation}
J.~Bergenholtz, W.~C. Poon, and M.~Fuchs,
\newblock Langmuir {\bf 19}, 4493 (2003).

\bibitem{chen2004microscopic}
Y.-L. Chen and K.~S. Schweizer,
\newblock J. Chem. Phys. {\bf 120}, 7212 (2004).

\bibitem{larson1999structure}
R.~G. Larson,
\newblock {\em The structure and rheology of complex fluids}, volume 150,
\newblock Oxford University Press (New York), 1999.

\bibitem{food-soft-materials2005}
R.~Mezzenga, P.~Schurtenberger, A.~Burbidge, and M.~Michel,
\newblock Nat. Mater. {\bf 4}, 729 (2005).

\bibitem{crop_protection2006}
M.~A. Faers, T.~H. Choudhury, B.~Lau, K.~McAllister, and P.~F. Luckham,
\newblock Colloids Surf. A Physicochem. Eng. Asp. {\bf 288}, 170 (2006).

\bibitem{darras2022}
A.~Darras et~al.,
\newblock Phys. Rev. Lett. {\bf 128}, 088101 (2022).

\bibitem{buscall1987consolidation}
R.~Buscall and L.~R. White,
\newblock J. Chem. Soc., Faraday Trans. 1 {\bf 83}, 873 (1987).

\bibitem{allain1995aggregation}
C.~Allain, M.~Cloitre, and M.~Wafra,
\newblock Phys. Rev. Lett. {\bf 74}, 1478 (1995).

\bibitem{allain2001systematic}
D.~Senis, L.~Gorre-Talini, and C.~Allain,
\newblock Euro. Phys. J. E {\bf 4}, 59 (2001).

\bibitem{starrs2002collapse}
L.~Starrs, W.~Poon, D.~Hibberd, and M.~Robins,
\newblock J. Phys. Cond. Mat. {\bf 14}, 2485 (2002).

\bibitem{allain2003rapid}
C.~Derec, D.~Senis, L.~Talini, and C.~Allain,
\newblock Physical Review E {\bf 67}, 062401 (2003).

\bibitem{weitz2005gravitational}
S.~Manley, J.~Skotheim, L.~Mahadevan, and D.~A. Weitz,
\newblock Phys. Rev. Lett. {\bf 94}, 218302 (2005).

\bibitem{bartlett2012sudden}
P.~Bartlett, L.~J. Teece, and M.~A. Faers,
\newblock Phys. Rev. E {\bf 85}, 021404 (2012).

\bibitem{zaccarelliHarich2016}
R.~Harich et~al.,
\newblock Soft Matter {\bf 12}, 4300 (2016).

\bibitem{padmanabhan2018gravitational}
P.~Padmanabhan and R.~Zia,
\newblock Soft Matter {\bf 14}, 3265 (2018).

\bibitem{zhou2018onset}
X.~Zhou,
\newblock (2018).

\bibitem{swan-furst2019}
K.~A. Whitaker et~al.,
\newblock Nat. Commun. {\bf 10}, 1 (2019).

\bibitem{zaccarelli2009colloidal}
E.~Zaccarelli and W.~C. Poon,
\newblock Proc. Nat. Acad. Sci. {\bf 106}, 15203 (2009).

\bibitem{carpineti1992spinodal}
M.~Carpineti and M.~Giglio,
\newblock Phys. Rev. Lett. {\bf 68}, 3327 (1992).

\bibitem{foffi2002evidence}
G.~Foffi et~al.,
\newblock Phys. Rev. E {\bf 65}, 050802 (2002).

\bibitem{zaccarelli2007colloidal}
E.~Zaccarelli,
\newblock J. Phys. Cond. Mat. {\bf 19}, 323101 (2007).

\bibitem{Royall2021}
C.~P. Royall, M.~A. Faers, S.~L. Fussell, and J.~E. Hallett,
\newblock J. Phys. Cond. Mat. {\bf 33}, 453002 (2021).

\bibitem{buzzaccaro2012equilibrium}
S.~Buzzaccaro, E.~Secchi, G.~Brambilla, R.~Piazza, and L.~Cipelletti,
\newblock Journal of Physics: Condensed Matter {\bf 24}, 284103 (2012).

\bibitem{happel1991low}
J.~Happel and H.~Brenner,
\newblock Low reynolds number hydrodynamics numbers, 1991.

\bibitem{landau1959fluid}
L.~Landau and E.~Lifshitz,
\newblock Section 92, problem {\bf 2} (1959).

\bibitem{carman1939permeability}
P.~C. Carman,
\newblock The Journal of Agricultural Science {\bf 29}, 262 (1939).

\bibitem{gray2004examination}
W.~Gray and C.~Miller,
\newblock Environmental science \& technology {\bf 38}, 5895 (2004).

\bibitem{carrillo2019darcy}
F.~J. Carrillo and I.~C. Bourg,
\newblock Water Resources Research {\bf 55}, 8096 (2019).

\bibitem{heijs1995numerical}
A.~W. Heijs and C.~P. Lowe,
\newblock Physical Review E {\bf 51}, 4346 (1995).

\bibitem{xu2008developing}
P.~Xu and B.~Yu,
\newblock Advances in water resources {\bf 31}, 74 (2008).

\bibitem{ozgumus2014determination}
T.~Ozgumus, M.~Mobedi, and U.~Ozkol,
\newblock Engineering Applications of Computational Fluid Mechanics {\bf 8}, 308 (2014).

\bibitem{ruiz2020tuning}
J.~Ruiz-Franco, F.~Camerin, N.~Gnan, and E.~Zaccarelli,
\newblock Physical Review Materials {\bf 4}, 045601 (2020).

\bibitem{koumakis2015tuning}
N.~Koumakis et~al.,
\newblock Soft Matter {\bf 11}, 4640 (2015).

\bibitem{gibaud2020rheoacoustic}
T.~Gibaud et~al.,
\newblock Physical Review X {\bf 10}, 011028 (2020).

\bibitem{Garbin-gel}
B.~Saint-Michel, G.~Petekidis, and V.~Garbin,
\newblock Soft Matter {\bf 18}, 2092 (2022).

\bibitem{torre2023structuring}
K.~W. Torre and J.~de~Graaf,
\newblock Soft Matter {\bf 19}, 2771 (2023).

\bibitem{torquato1995mean}
S.~Torquato,
\newblock Physical review letters {\bf 74}, 2156 (1995).

\bibitem{ribeiro2016helmholtz}
P.~C. Ribeiro, H.~F. de~Campos~Velho, and H.~Lopes,
\newblock Computers \& Graphics {\bf 55}, 80 (2016).

\bibitem{klaseboer2019helmholtz}
E.~Klaseboer, Q.~Sun, and D.~Y. Chan,
\newblock Journal of Elasticity {\bf 137}, 83 (2019).

\bibitem{glotzl2023helmholtz}
E.~Gl{\"o}tzl and O.~Richters,
\newblock Journal of Mathematical Analysis and Applications {\bf 525}, 127138 (2023).

\bibitem{evans2002emergence}
R.~M.~L. Evans and L.~Starrs,
\newblock Journal of Physics: Condensed Matter {\bf 14}, 2507 (2002).

\bibitem{john2024early}
T.~John, L.~Kaestner, C.~Wagner, and A.~Darras,
\newblock PNAS nexus {\bf 3}, pgad416 (2024).

\bibitem{varga2018modelling}
Z.~Varga, J.~L. Hofmann, and J.~W. Swan,
\newblock Journal of Fluid Mechanics {\bf 856}, 1014 (2018).

\bibitem{de2023hydrodynamic}
J.~De~Graaf, K.~W. Torre, W.~C. Poon, and M.~Hermes,
\newblock Physical Review E {\bf 107}, 034608 (2023).

\bibitem{razali2017effects}
A.~Razali et~al.,
\newblock Soft Matter {\bf 13}, 3230 (2017).

\bibitem{meeker1998low}
S.~P. Meeker,
\newblock KB thesis scanning project 2015  (1998).

\bibitem{versteeg1995computational}
H.~Versteeg and W.~Malalasekera,
\newblock The finite volume method , 1 (1995).

\end{thebibliography}

\appendix

\section{\label{sec:appen}Numerical Scheme}

In this appendix, we provide the details of the numerical scheme used to solve our model in 1D. For non-homogeneous density profiles, by combining equations~\eqref{eq:ome} with~\eqref{eq:qseqn}, we obtain a non-homogeneous linear differential equation for the gel stress $\sigma_c(z)$
\begin{align}
    \sigma_c &= \lambda(\phi) \ \partial_z \left [ (1-\phi) K(\phi) \left (-v_g \phi + \frac{\sigma^2}{\mu} \partial_z \sigma_c \right ) \right ] ; \\
    &= A(\phi)  + B(\phi)  \partial_z \sigma_c + C(\phi) \partial^2_z \sigma_c ,
\end{align}
where the non-constant coefficients are defined as
\begin{align}
    A(\phi) &= - \partial_z  \phi \ \lambda(\phi) \ \partial_{\phi} \left [ \phi (1-\phi) K(\phi) \right ] v_g ; \\
    B(\phi) &= \partial_z \phi \ \lambda(\phi)  \ \partial_{\phi} \left [ (1-\phi) K(\phi) \right ] \mu^{-1} \sigma^2 ; \\
    C(\phi) &= \lambda(\phi)(1-\phi) K(\phi)  \mu^{-1} \sigma^2,
\end{align}
and the boundary conditions are given by
\begin{align}
    \left. \partial_z \sigma_c  \right |_{z=0,H} = \Delta \rho g \ \phi|_{z=0,H}.   
\end{align}
The system height $H$ is subdivided in $N+1$ intervals of size $\delta z$, and all fields are represented on this grid using a superscript indicating their evaluation positions. We employ central second-order and first-order finite difference schemes in the bulk and at the boundaries, respectively, to derive the following finite-difference equations for $\sigma_c$:
\begin{align}
    \left . \sigma^i_c \right |_{0<i<N} &= \frac{A^i+ \frac{ B^i}{2\delta z }\left ( \sigma_c^{i+1} - \sigma_c^{i-1} \right )+ \frac{C^i}{\delta z^2}(\sigma_c^{i+1} + \sigma_c^{i-1})} {1 + \frac{2C^i}{ \delta z^2}} ; \\ 
    \left . \sigma^i_c \right |_{i=0} &= -\Delta \rho g \ \phi^0 + \sigma^1_c ;\\
    \left . \sigma^{i}_c \right |_{i=N} &= \Delta \rho g \ \phi^N - \sigma^{N-1}_c .
\end{align}
These equations are iteratively solved to update the stress values until convergence is reached, with the tolerance set to $0.001$\%.

Once the numerical solution for the stress is computed, it is divided by the dilatational viscosity $\lambda$, which is computed using the volume fraction of the current time step, to obtain $\partial_z v_c(z,t)$. Next, we integrate this result to find the colloid velocity field:
\begin{align}    
    \int_0^z dz' \ \partial_{z'} v_c(z',t)   =  v_c(z,t),
\end{align}
where the boundary velocity term $v_c(0,t)$ is simplified using mass conservation and the fact that the system is isolated. An approximation of the above integral is computed using the trapezoidal rule, resulting in the discretized colloid velocity:
\begin{align}    
    \left.  v_c^i \right |_{0<i<N} &= \frac{\delta z}{2} \sum_{j=1}^i \left ( \frac{\sigma_c^{j+1}}{\lambda^{j+1}} + \frac{\sigma_c^{j-1}}{\lambda^{j-1}} \right ) ; \\
    \left. v_c^i \right |_{i=0} &= \left. v_c^i \right |_{i=N} = 0.
\end{align}

Now that the colloid velocity field is known, we can advance the colloid volume fraction one time step forward using equation~\eqref{volumebalance}. This is solved numerically using an upwind finite-volume method~\cite{versteeg1995computational}. Standard finite-difference schemes would fail to converge due to the strong convective nature of equation~\eqref{volumebalance}, given the absence of explicit diffusive terms. Thus, the change in volume fraction is approximately computed at the center of each of the $N+1$ space intervals using the biased fluxes
\begin{align}
J_{\text{up}}^k &= \begin{cases}
        \phi^{k} v_c^{i+1} & v_c^{i+1}>0 \\
        \phi^{k+1}  v_c^{i+1} & v_c^{i+1}<0 
    \end{cases} ; \\
J_{\text{down}}^k &= \begin{cases}
        \phi^{k-1} v_c^{i} & \ \ \ v_c^i>0 \\
        \phi^{k} v_c^{i} & \ \ \ v_c^i<0 
    \end{cases} ,
\end{align}
where we have introduced the indices $k = i + 1/2$ to relate the velocity field, computed at the $N+1$ nodes, with the volume fraction field, computed at the $N$ centers of the intervals formed by consecutive nodes. The instantaneous change in volume fraction at a given time step can then be computed using
\begin{align}    
      \partial_t \phi^{k}(t) &= - J_{\text{up}}^k + J_{\text{down}}^k,
\end{align}
and the updated values of $\phi$ are obtained using a first-order forward Euler scheme:
\begin{align}    
\phi^{k}(t+\delta t) = \delta t \ \partial_t \phi^{k}(t).
\end{align}

\end{document}